\documentclass[pra,print,superscriptaddress,twocolumn]{revtex4}
\usepackage{amssymb}
\usepackage{amsmath}
\usepackage{epsfig}
\usepackage{color}
\usepackage{graphics, graphicx}
\usepackage{bbold}
\usepackage{psfrag}
\usepackage{mathcomp}
\usepackage{subfigure}
\usepackage{verbatim}
\usepackage{color}
\usepackage[colorlinks,citecolor=blue]{hyperref}

\begin{document}

\title{Quantum phases of the biased two-chain-coupled Bose-Hubbard Ladder}
\author{Jingtao Fan}
\affiliation{State Key Laboratory of Quantum Optics and Quantum Optics Devices, Institute
of Laser Spectroscopy, Shanxi University, Taiyuan 030006, China}
\affiliation{Collaborative Innovation Center of Extreme Optics, Shanxi University,
Taiyuan 030006, China}
\author{Xiaofan Zhou}
\thanks{zhouxiaofan@sxu.edu.cn}
\affiliation{State Key Laboratory of Quantum Optics and Quantum Optics Devices, Institute
of Laser Spectroscopy, Shanxi University, Taiyuan 030006, China}
\affiliation{Collaborative Innovation Center of Extreme Optics, Shanxi University,
Taiyuan 030006, China}
\author{Suotang Jia}
\affiliation{State Key Laboratory of Quantum Optics and Quantum Optics Devices, Institute
of Laser Spectroscopy, Shanxi University, Taiyuan 030006, China}
\affiliation{Collaborative Innovation Center of Extreme Optics, Shanxi University,
Taiyuan 030006, China}

\begin{abstract}
We investigate the quantum phases of bosons in a two-chain-coupled ladder.
This bosonic ladder is generally in a biased configuration, meaning that the
two chains of the ladder can have dramatically different on-site
interactions and potential energies. Adopting the numerical
density-matrix renormalization-group method, we analyze the phase transitions
in various parameter spaces. We find signatures of both
insulating-to-superfluid and superfluid-to-insulating quantum\ phase
transitions as the interchain tunnelling is increased. Interestingly,
tunning the interaction to some intermediate values, the system can exhibit
a reentrant quantum phase transition between insulating and superfluid
phases. We show that for infinite interaction bias, the model is amenable to
some analytical treatments, whose prediction about the phase boundary is
in great agreement with the numerical results. We finally clarify some
critical parameters which separate the system into regimes with distinct
phase behaviours, and briefly compare typical properties of the biased and
unbiased bosonic ladder systems. Our work enriches the Bose-Hubbard physics.
\end{abstract}

\maketitle

\section{Introduction}

Strongly correlated bosons, especially those moving in the periodic
potentials, have always\ been the research interest for both
experimentalists and theorists, as they are related to a variety of quantum
phenomenon \cite{Boson1,Boson2}. The simplest model describing such systems
is the Bose-Hubbard (BH) model, which incorporates the contributions from
the kinetic energy of individual atoms and the repulsive interactions
between them \cite{BHM1,BHM2,BHM3,BHM4,BHM5,BHM6}. Although originally
developed in the context of $^{4}$He liquid \cite{BHM1}, it has been
demonstrated that the BH model can be feasibly implemented with ultracold
atoms trapped in optical lattices \cite{Book1,OL1,OL2,OL3}. Utilizing the
unprecedented degree of controllability of the laser fields, all the
characteristic parameters of the BH model can be tuned in the optical
lattice with high precision \cite{OL2,OL3}. Relying on this, the quantum
phase transition from a superfluid (SF) to a Mott insulator (MI), which is
the most important prediction of the BH model, has been experimentally
realized in one \cite{OL4}, two \cite{OL5} and three dimensions \cite{OL6}.
Since then, lots of related studies have been performed on the extensions of
the BH model, by considering, for example, diverse forms of interactions
\cite{ExtendBH1,ExtendBH2,ExtendBH3} and gauge fields \cite{ExtendBH4}.
These extensions stimulate the development of new directions bridging
condensed matter physics, stochastic physics and quantum optics.

In this context, the bosons confined to low-dimensional lattices merit
special attention, since the correlations built up in these systems are
considerably enhanced by the interactions between atoms \cite{OneD}. Among
various low-dimensional lattice models, the two-chain-coupled BH ladder is
of particular importance \cite{BHLtheory1,BHLtheory2}, since it serves as an
intermediate geometry intervening the one and two lattice systems \cite%
{Ladder}. This provides beneficial insights about the characteristics of the
SF-to-MI transition in going from one to two dimensions. As a matter of
fact, the BH ladder has been experimentally simulated in different
artificial systems \cite{BHLexperiment1,BHLexperiment2,BHLexperiment3},
stimulating immense interests of research towards various aspects of this
model, such as the chiral currents \cite{Current1,Current2,Current3,Current4,Current5,Current6}, the
quantum magnetism \cite{Magnetism1,Magnetism2,Magnetism3,Magnetism4,Magnetism5}, and
the topological states \cite{Topo1,Topo2,Topo3,Topo4,Topo5,Topo6}. The BH ladders
considered in these studies, however, are mostly limited to the symmetric
case where both the on-site interactions and potential energies are
identical for the two chains. Notice that general ladder systems should also
involve the configurations where the two chains have distinctly different
system parameters. Actually, letting the two chains of the ladder
asymmetric, with respect to either the interactions or potential energies,
may impose major impacts on various properties of the system \cite%
{Biased1,Biased2,Biased3,Biased4,Biased5}. More importantly, this
\textquotedblleft biased\textquotedblright\ ladder structure also underlies
the physics behind a large class of systems, such as the dressed dipolar
molecules \cite{Dipolar1,Dipolar2} or Rydberg gases \cite%
{Rydberg1,Rydberg2,Rydberg3} in optical lattices and the low-dimensional
magnetic materials under external fields \cite{Material1,Material2}.

In this work, we investigate the ground-state properties of a biased bosonic
ladder at half filling, using state-of-the-art density-matrix
renormalization-group (DMRG) numerical methods \cite{dmrg1,dmrg2}. By saying \textquotedblleft
biased\textquotedblright , we mean that the two chains constituting the
whole ladder can have dramatically different on-site interactions and
potential energies. We first provide an analysis of the quantum phases in
the limit of infinite interaction bias, where the on-site interaction is
infinite for one chain and finite for the other. It is found that, as the
interchain tunnelling is increased, either the MI-to-SF or the SF-to-MI
quantum\ phase transition can occur depending on the value of interactions.
More interestingly, tuning the finite interaction to some intermediate
values, the system may even exhibit a reentrant quantum phase transition
between MI and SF. By mapping the finite interaction into an effective
canonical Kerr nonlinear form, we analytically derive the phase boundary
between MI and SF, which agrees well with the numerical results. With the
knowledge of the system under infinite interaction bias, we then discuss the
more general parameter regime where the interactions of both chains of the
ladder are finite. We map out the ground-state phase diagrams in various
parameter spaces, and characterize several critical parameters which
separate the system into regimes with distinct phase behaviours. Finally, we
briefly compare the typical properties of the biased and unbiased bosonic
ladder systems.

\section{Model and method}

\label{sec:system}

\begin{figure}[tp]
\includegraphics[width=8.0cm]{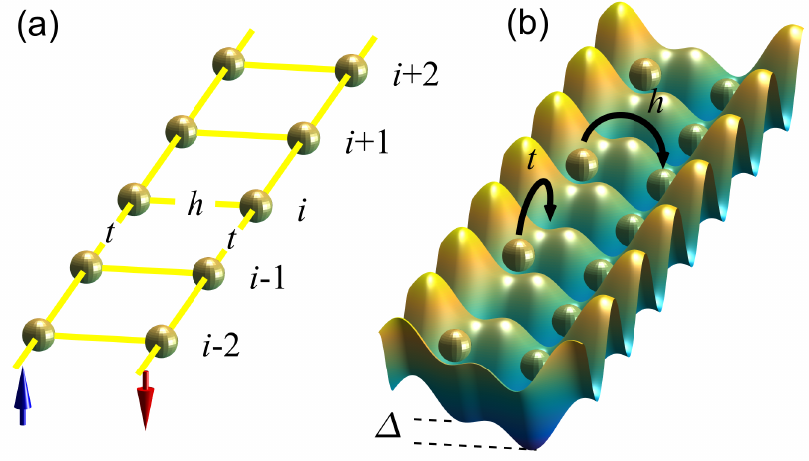}
\caption{(a) Schematic picture of the ladder system. The two chains
constituting the ladder are designated as spin-up and spin-down,
respectively. The interspin and intraspin hopping amplitudes are denoted,
respectively, as $t$ and $h$. (b) Possible implementation of the bosonic
ladder in optical superlattices. The optical double well is generally tilted
by an energy difference $\Delta $. The boson tunnelling rates along
different directions simulate the hopping rates $t$ and $h$.}
\label{Fig1}
\end{figure}

As illustrated in Fig.~\ref{Fig1}(a), the system in consideration is a
bosonic ladder with two coupled chains, which we denote as spin-up and
spin-down, respectively. The interspin tunnelling is allowed along the rung.
We assume the bosonic ladder is typically biased, i.e., atoms with different
spins experience different potential energies and local interactions. Such a
scenario can be effectively engineered in spin-dependent optical lattices
\cite{SOL1} or optical superlattices \cite{Current3,DOL1}, where the spin
index distinguishing different chains can be represented by either\ the
hyperfine sublevels or optical wells, according to different experiment
implementations [see Fig.~\ref{Fig1}(b) for illustration]. The Hamiltonian
describing this system reads%
\begin{eqnarray}
\hat{H} \!\! &=& \!\! -t\sum_{\left\langle i,j\right\rangle ,\sigma }\hat{b}_{i,\sigma }^{\dag
}\hat{b}_{j,\sigma }-h\sum_{j}(\hat{b}_{j,\uparrow }^{\dag }\hat{b}_{j,\downarrow }+\text{H.c.}%
)  \notag \\
&& \!\! +\Delta \sum_{j}(\hat{n}_{j,\uparrow }-\hat{n}_{j,\downarrow })+\sum_{j,\sigma }\frac{%
U_{\sigma }}{2}\hat{n}_{j,\sigma }\left( \hat{n}_{j,\sigma }-1\right)  \label{BBH}
\end{eqnarray}%
where the field operator $\hat{b}_{j,\sigma }$ ($\hat{b}_{j,\sigma }^{\dag }$)
annihilates (creates) a bosonic atom with spin $\sigma $ (=$\uparrow
,\downarrow $) at the lattice site $j$. While\ atoms with the same spin can
hop between adjacent sites $\left\langle i,j\right\rangle $ with the
intraspin hopping rate $t$, an interspin field along the rung of the ladder
couples atoms with different spins at rate $h$. The energy bias $\Delta $,
which we assume to be positive in this work, tends to polarize the atoms
along the rung of the ladder and $U_{\sigma }$ denotes the on-site repulsive
interaction of atoms with spin $\sigma $ ($=\uparrow ,\downarrow $). In this
work, we focus on the commensurate ladder with total atomic density to be $%
\rho =N/2L=1/2$. Here, $N=N_{\uparrow }+N_{\downarrow }$ is the total number
of bosons on the two-chain ladder, each of which has length $L$. This
amounts to setting the total system size to be $2L$. In the following
discussion, we set the energy scale by taking $t=1$, and we also take $%
\Delta =10$ unless otherwise specified.\

The Hamiltonian~(\ref{BBH}) can be viewed as a natural extension of the
single-component BH model incorporating the spin degree of freedom, which is
controlled by both the transverse and longitude magnetic fields. Without the
interspin tunnelling $h$, the ladder decouples and reduces to two
independent BH chains. A finite energy bias $\Delta $ ($>0$) then polarizes
the bosons to the spin-down chain with commensurability of one boson per
site, leaving the spin-up chain empty. In this case, the physics is entirely
governed by the one-dimensional BH model, which has been extensively
explored \cite{BHM1,BHM2,BHM3,BHM4,BHM5,BHM6}. With a non-zero interspin
tunnelling $h$, however, the two BH chains are coupled together, meaning
that the characteristic parameters of each individual chain may impact the
ground-state properties of the composite system in a collective manner. This
becomes especially interesting if the on-site interactions of the two chains
are tuned quite different. Without loss of generality, let us assume $%
U_{\uparrow }>U_{\downarrow }$ and take a preanalysis on the behavior of the
BH chain with spin-down. In this case, the on-site interaction may
effectively enhanced through high-order tunnelling process triggered by $h$
\cite{Rydberg4}, favoring the formation of MI, whereas the particle filling
factor may deviate from unity at the same time, which in turn promotes the
SF character. The seemingly opposite tendency of the ground-state property
makes the roles of the parameters in the biased bosonic ladder less
intuitive and to be quantitatively clarified.

The MI and SF phases can be directly identified by calculating the charge gap $\delta_{L}$, defined as the difference between
the energies needed to add and remove one particle from the system, i.e.,%
\begin{equation}
\delta _{L}=\mu _{L}^{+}-\mu _{L}^{-},
\end{equation}%
\begin{equation}
\mu _{L}^{+}=E_{L}(N+1)-E_{L}(N),
\end{equation}%
\begin{equation}
\mu _{L}^{-}=E_{L}(N)-E_{L}(N-1),
\end{equation}%
where $E_{L}(N)$ is the the ground-state energy for $L$ sites and $N$
particles, and the chemical potentials $\mu _{L}^{+}$ and $\mu _{L}^{-}$
characterize the energy cost to add and remove one particle, respectively.
The insulating phase is signaled by the opening up of $\delta _{L}$ in the
thermodynamical limit $N,L \rightarrow \infty $ with fixed density $\rho $,
consisting with\ the zero compressibility $\kappa $ ($=\partial \rho
/\partial \mu $) of an insulator \cite{CF1}. In the SF phase, however, the
charge gap $\delta _{L}$ closes and the system becomes compressible in the
thermodynamical limit. Since $\delta _{L}$ keeps finite for any finite
systems, in order to pinpoint the MI-to-SF transition in the parameter space,
we should extrapolate to the $L \rightarrow \infty $ limit by utilizing the
standard finite-size scaling. Here, we perform state-of-the-art DMRG
calculations to compute the many-body ground state of the system, with which
various physical observable can be obtained. In our numerical simulations,
we set the cutoff of the single-site atom number as $n_{\text{cutoff}}=6$.
We set lattice size up to $L=40$,
for which we retain 600 truncated states per DMRG block and perform 20
sweeps with a maximum truncation error of $\sim 10^{-9}$.

\section{Results}

\label{sec:results}

In the following, we systematically study the ground-state properties of the
biased bosonic ladder. Before providing the results of general parameters,
we first put the interaction bias to infinity, i.e., we consider the bosonic
ladder consisting of one chain with infinite on-site interaction and the
other with finite on-site interaction. The physics in this limit serves as a
beneficial starting point to understand the essential mechanism behind which
various phase behaviours occur.

\subsection{Infinite interaction bias: $U_{\uparrow }\!-\!U_{\downarrow } = \infty
$}

\label{sec:results:infinite}

Without loss of generality, we fix the on-site interaction of the spin-up BH
chain to be infinity, $U_{\uparrow }\rightarrow \infty $, and that of the spin-down
chain to be finite. This amounts to imposing a hard-core constraint on each
site of the spin-up chain, where only one boson is allowed to occupy.

Before showing the full phase diagram, we can gain some useful insights into
the system by inspecting certain limit parameters. As mentioned in Sec.~\ref%
{sec:system}, the simplest limit is the zero interspin tunnelling $h=0$,
under which the ground state is fully described by the one-dimensional BH
model with unit filling. It is well known that the BH model with unit filling shows a SF-to-MI
transition at $U_{\text{c}}/t\sim 3.3$ \cite{BHM4,BHM5,BHM6}. The physics becomes
richer when $h$ is turned on. In this case, if we further set the on-site
interaction $U_{\downarrow }$ to be zero, the system closely resembles that
of two-level atoms inside cavity arrays for which the
Jaynes-Cummings-Hubbard (JCH) model works \cite{JCH1,JCH2,JCH3,JCH4}. This
can be seen clearly if we map the field operators of the hardcore bosons to
those of quantum spins by $\hat{b}_{j,\uparrow }\longrightarrow \sigma _{-}$ and $%
\hat{b}_{j,\uparrow }^{\dag }\longrightarrow \sigma _{+}$, where $\sigma _{-}$ and
$\sigma _{+}$ are spin-1/2 lowering and raising operators, respectively.
Therefore, the Hamiltonian describing the tunnelling process between the two
BH chains becomes Jaynes-Cummings type, in which the spin-down bosons act as
interaction-free photons bridging adjacent lattice sites. It follows
directly from the JCH physics that, by increasing $h$, the spin-down bosons
behave more localized and consequently undergoes a phase transition from SF
to MI at some critical tunnelling strength $h_{\text{c}}$ \cite{JCH1,JCH2}.

Anther interesting limit which is opposite to the JCH regime is $%
U_{\downarrow }\rightarrow \infty $, implying a hard-core constraint on both chains. It
has been analytically demonstrated that, for the symmetric hard-core BH
ladder with $\Delta =0$, the critical tunneling strength $h_{\text{c}}$ decreases
down to zero \cite{BHL1}. As a finite energy bias $\Delta $ usually tends to
increase the band gap, we do not expect the critical tunneling strength $%
h_{\text{c}}$ to be shifted away from that of the symmetric case (i.e., $\Delta =0$%
).

\begin{figure}[tp]
\includegraphics[width=8.5cm]{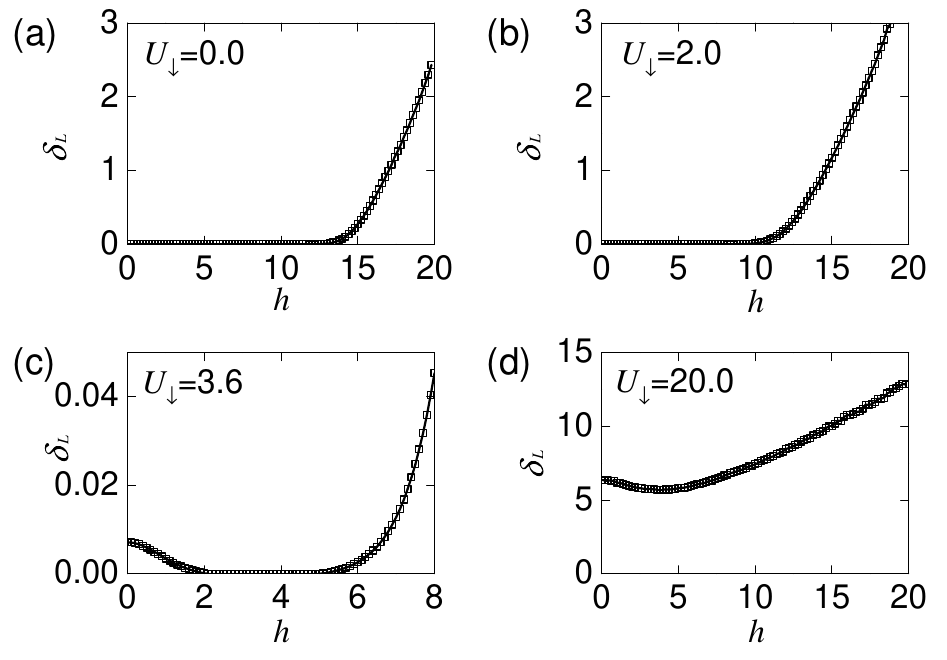}
\caption{The charge gap $\protect\delta _{L}$, extroplated to $L \rightarrow
\infty $, as a function of $h$ with (a) $U_{\downarrow }=0.0$, (b) $%
U_{\downarrow }=2.0$, (c) $U_{\downarrow }=3.6$ and (d) $U_{\downarrow }=20.0$.
The other parameters are $\Delta =10$ and $U_{\uparrow} = \infty$. }
\label{Fig2}
\end{figure}

\begin{figure}[tp]
\includegraphics[width=8.5cm]{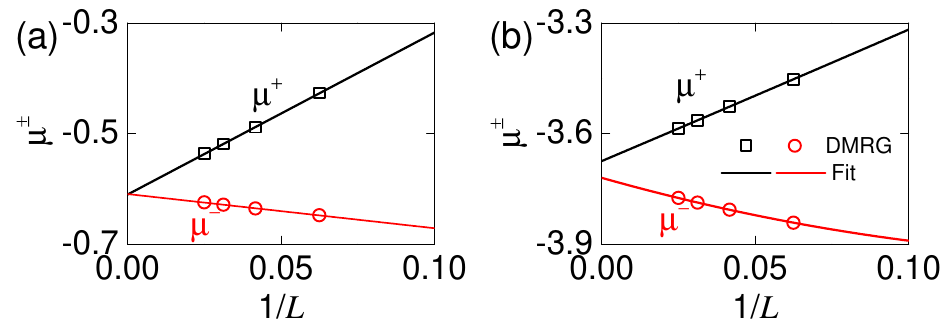}
\caption{The finite-size scaling of chemical potentials $\protect\mu %
_{L}^{+} $ and $\protect\mu _{L}^{-}$\ for (a) $h=3.0$ and (b) $h=8.0$.\ The
other parameters are $\Delta =10$, $U_{\downarrow } = 3.6$ and $U_{\uparrow } = \infty $. }
\label{Fig3}
\end{figure}

We now start to show the numerical results of general parameters. In order
to clarify the effect of the interspin tunnelling on the SF-to-MI transition,
we vary $h$ from zero to some large value and calculate the corresponding
charge gap $\delta _{L}$, which ought to be extrapolated to the $%
L \rightarrow \infty $ limit. From the knowledge of the one-dimensional BH
model, the system stays at the SF phase (MI phase) for $U_{\downarrow
}\lesssim 3.3$ ($U_{\downarrow }\gtrsim 3.3$) and $h=0$, and potential phase
transitions can take place when increasing $h$. Figure \ref{Fig2}(a) plots
the charge gap $\delta _{L}$ as a function of $h$ for $U_{\downarrow }=0$.
It can be seen that, starting from zero, the charge gap gradually opens up
when $h$ exceeds the critical value $h_{\text{c}}\approx 12.5$, evidencing a SF-to-MI
transition, as can be inferred from the JCH physics \cite{JCH2}. The
critical tunnelling strength $h_{\text{c}}$ decreases as we increase the on-site
interaction $U_{\downarrow }$, as shown in Fig.~\ref{Fig2}(b). In Fig.~\ref%
{Fig2}(d), we exemplify another limit where the on-site interaction is
considerably strong by setting $U_{\downarrow }=20$. It can be found that,
in this case, the gap $\delta _{L}$ keeps open irrespective of the value of $%
h$, meaning the system remains a MI. Something interesting happens when the
on-site interaction $U_{\downarrow }$ is tuned slightly larger than $U_{\text{c}}$.
As illustrated in Fig. \ref{Fig2}(c), we plot $\delta _{L}$ versus $h$ for $%
U_{\downarrow }=3.6$. With the increase of $h$, the gap first closes at $%
h\approx 2.0$ and then reopens at $h\approx 5.1$, indicating that the phase
transition appears twice. That is, the system starts from the MI, and
subsequently traverses the SF phase, and ending up in the MI eventually.
Figures~\ref{Fig3}(a) and \ref{Fig3}(b) show finite-size scaling of the DMRG
data of the charge gap, by linear and quadratic fittings, for two
representative points located in the SF and MI phases, respectively. This
reentrant MI phase transition induced by the interspin tunnelling strength $h
$, does not exist in the symmetric bosonic ladder with $U_{\uparrow
}=U_{\downarrow }$ and $\Delta =0$, and is thus exclusive for the biased
ladder here.

\begin{figure}[tp]
\includegraphics[width=8.0cm]{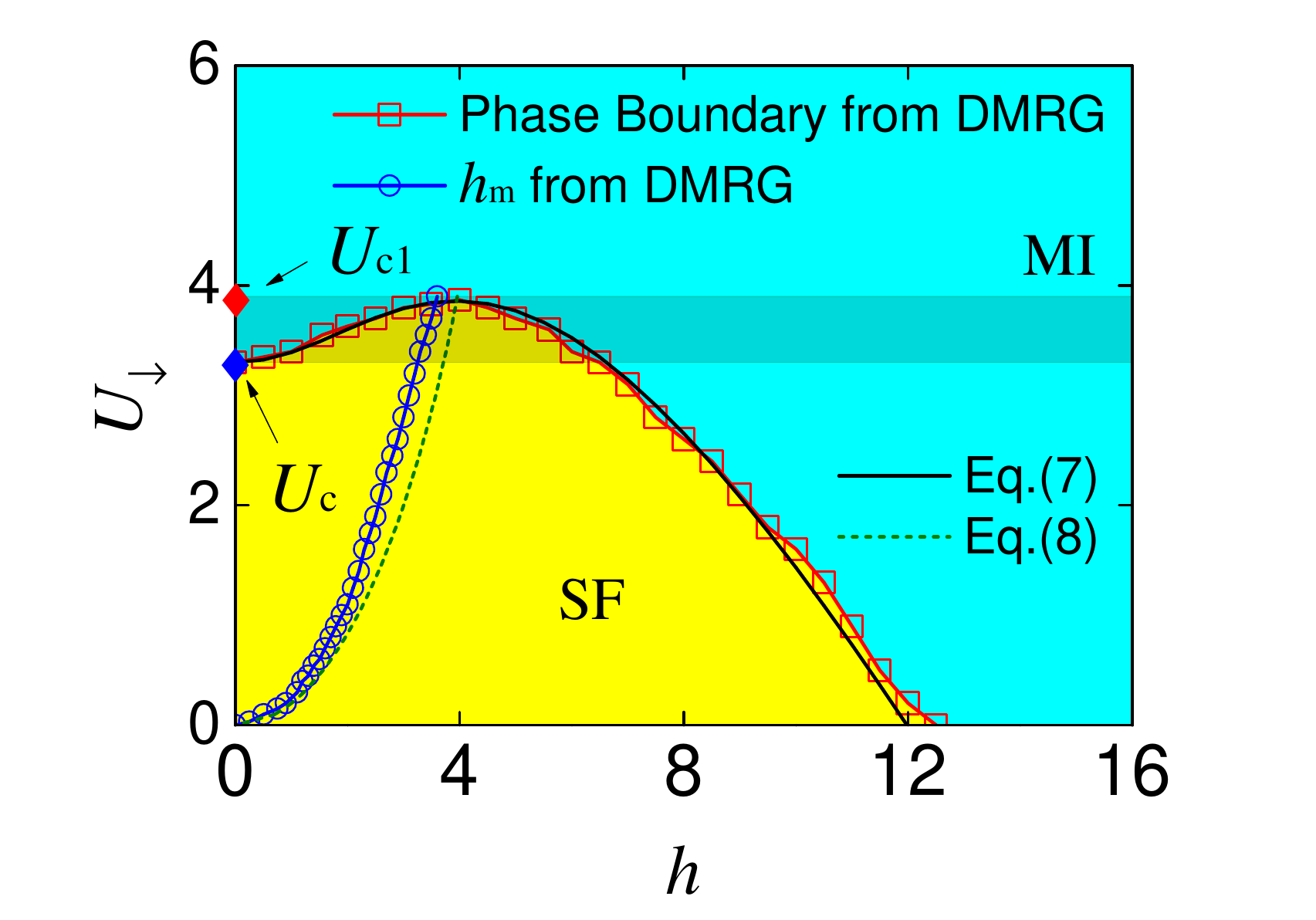}
\caption{The phase diagram in the $U_{\downarrow }-h$ plane for $\Delta =10$
and $U_{\uparrow } = \infty $.  The red solid line with square symbol (black solid line) denotes The MI-SF phase boundary obtained by the DMRG calculation [Eq.~(\protect\ref{Boundary})]. The values of $h_{\text{m}}$ for each $U_{\downarrow }$ is also pinpointed in the phase diagram by the blue solid line with circle symbol. For comparation, location of the minimum of $U_{\text{eff}}$, determined by Eq.~(\protect\ref{Umin}), is plotted by the green dashed line. The shaded area, bounded by critical interactions $U_{\text{c}}\approx 3.3$ and $U_{c1}\approx 3.9$, characterizes the parameter region where the MI-to-SF-to-MI transition can occur.}
\label{Fig4}
\end{figure}

With the understanding above, we map out the phase diagram in the $%
U_{\downarrow }-h$ plane in Fig.~\ref{Fig4}. The phase boundary has been
extrapolated to the $L \rightarrow \infty $ limit by the finite-size scaling.
It is to be seen clearly that, while increasing the on-site interaction $%
U_{\downarrow }$ always drives the system to the MI phase, the role of the
interspin tunnelling $h$ can be somehow opposite, i.e., it can trigger both
the MI and SF phases, depending on the value of $U_{\downarrow }$. Notice
that the MI in the BH model is essentially stabilized by the direct
interaction between bosons, whereas bosons with different spins are dressed
together here forming composite polaritons. We therefore expect that some
effective interaction between polaritons, which plays the key role in
inducing different behaviours of the phase transition, may emerge.

\begin{figure}[tp]
\includegraphics[width=8.0cm]{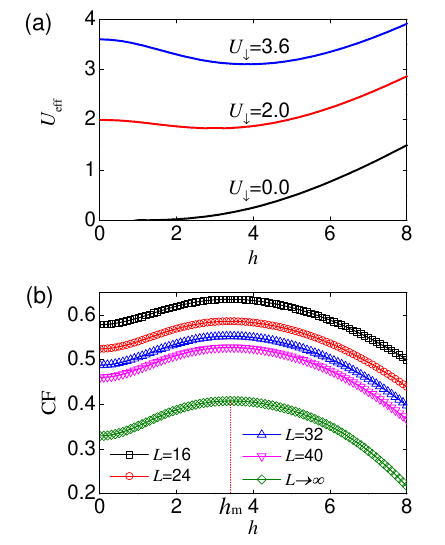}
\caption{(a) The effective Hubbard interaction $U_{\text{eff}}$ as a
function of $h$ for different $U_{\downarrow }$ with $\Delta =10$ and $%
U_{\uparrow } = \infty $. (b) The
condensate fraction calculated for different system sizes. Note that the
result of $L\rightarrow \infty $ is obtained by extrapolation using the finite-size
scaling. $h_{\text{m}}$ specifies the location of the maximum of the condensate fraction. The other parameters are $\Delta =10$, $U_{\downarrow }=3.6$ and $%
U_{\uparrow } = \infty $.}
\label{Fig5}
\end{figure}

To see this clearly, we map the local contribution of the Hamiltonian~(\ref%
{BBH}) into an effective kerr nonlinearity by a simple energy mismatch
argument. As detailed in Appendix A, the local energy of Hamiltonian~(\ref%
{BBH}) consists of two plaritonic modes whose eigenenergies are%
\begin{eqnarray}
\omega _{n}^{\pm } &=&\frac{U_{\downarrow }}{2}n(n-2)+\frac{1}{2}(\Delta
+U_{\downarrow })  \notag \\
&&\pm \frac{1}{2}\sqrt{(nU_{\downarrow }-U_{\downarrow }-\Delta )^{2}+4nh^{2}%
}  \label{PE}
\end{eqnarray}%
where $n$ is the excitation number. Since we are only interested in the
low-energy physics, the focus in the following will be on the lower branch $%
\omega _{n}^{-}$. We define the effective Hubbard interaction $U_{\text{eff}%
} $ as the energy cost incurred by forming a two-particle plaritonic
excitation (with energy $\omega _{2}^{-}$) from two single-particle
plaritonic excitation (with energy 2$\omega _{1}^{-}$) in neighboring
lattice sites \cite{JCH3,JCH4}, i.e.,%
\begin{eqnarray}
U_{\text{eff}} &=&\omega _{2}^{-}-2\omega _{1}^{-}  \notag \\
&=&\frac{1}{2}(U_{\downarrow }-\Delta )+\sqrt{\Delta ^{2}+4h^{2}}  \notag \\
&&-\frac{\sqrt{(U_{\downarrow }-\Delta )^{2}+8h^{2}}}{2}  \label{EffU}
\end{eqnarray}%
With this understanding, we can obtain an analytical expression of the phase
boundary between the MI and SF phases by equating the effective interaction $%
U_{\text{eff}}$ with the critical interaction strength of the BH model with
unit filling, namely
\begin{equation}
U_{\text{eff}}=U_{\text{c}}\approx 3.3  \label{Boundary}
\end{equation}%
As shown in Fig.~\ref{Fig4}, the curve defined by Eq.~(\ref{Boundary}),
agrees well with the numerical results obtained by the DMRG calculation. It
should be emphasized that, in deriving Eq.~(\ref{EffU}), we have implicitly
assumed that the ground-state property of the whole lattice system is
mainly governed by its low-energy local physics. This requires that (i) the
energy scale owned by each local lattice sites is considerably larger than
the kinetic energy of bosons, namely at least $\Delta \gg 1$ or $h\gg 1$,
and (ii) the density fluctuations are weak enough so that only the
lowest-lying excitations of individual lattice sites need to be taken into
consideration. This guarantees the effectiveness of Eq.~(\ref{Boundary}) in
predicting the MI-to-SF phase boundary, since the density fluctuations are
extremely suppressed in the MI. Eq.~(\ref{EffU}) provides further guidance
to the driving force inducing different phase transitions. An interesting
finding is that $U_{\text{eff}}$ exhibits nonmonotonic behaviour as $h$
increases from zero. As illustrated in  Fig.~\ref{Fig5}(a), with the increase
of $h$, the effective interaction $U_{\text{eff}}$ decreases first to a
minimum and then increases monotonically [see blue line], which explains the MI-to-SF-to-MI
transition found in Fig.~\ref{Fig4}. The location of the minimum of $U_{%
\text{eff}}$ can be easily deduced by requiring $\partial U_{\text{eff}%
}/\partial h=0$, yielding a trivial solution $h=0$ and a nontrivial solution,%
\begin{equation}
h=\frac{\sqrt{\Delta ^{2}-(\Delta -U_{\downarrow })^{2}}}{2}.  \label{Umin}
\end{equation}%
It is straightforward to show that Eq.~(\ref{Umin}), within its range of
values, minimizes $U_{\text{eff}}$. The curve obtained from Eq.~(\ref{Umin})
is depicted\ in Fig.~\ref{Fig4}. Notice that, whereas $U_{\text{eff}}$ is
minimized by $h=0$\ when $U_{\downarrow }=0$, consistent with the JCH
physics \cite{JCH1,JCH2}, a nonzero $U_{\downarrow }$\ shifts the location
of the interaction minimum (i.e., $h=0$) to some finite value. Within this
picture, an upper bound of $U_{\downarrow }$, beyond which no SF phase would
exist, can be obtained. This is immediately achieved by substituting Eq. (%
\ref{Umin}) into Eq.~(\ref{Boundary}), which is then solved by
\begin{equation}
U_{\downarrow }=U_{\text{c1}}\equiv \Delta +U_{\text{c}}-\sqrt{\Delta ^{2}-U_{\text{c}}^{2}}.
\end{equation}
It becomes clear that, the parameter region of $U_{\downarrow }$ within
which the MI-to-SF-to-MI transition can occur is $U_{\text{c}}<$ $U_{\downarrow }<U_{\text{c1}}$%
.

An experimental measurable quantity that is able to mirror the effective
interaction is the condensate fraction (CF), defined as the number of bosons in
the condensate with respect to the total number of bosons \cite{CF1,CF2}. It
has been shown that the condensate fraction of Bose gases monotonically
decreases as the local interaction increases \cite{CF1}. For the bosonic
ladder considered here, the CF is defined as the largest eigenvalue of the
matrix $\left\langle \hat{b}_{i,\sigma }^{\dag }\hat{b}_{j,\sigma'}\right\rangle $
divided by the total number of bosons \cite{CF2}. Figure~\ref{Fig5}(b) shows
the CF as a function of $h$ for $U_{\downarrow }=3.6$ and different system
sizes. It is demonstrated that the CF increases first, reaching its maximum
at $h\approx 3.4$, and then decreases. The location of\ the maximum of CF,
designated as $h_{\text{m}}$, depends sensitively on the value of $%
U_{\downarrow }$. As shown in Fig.~\ref{Fig4}, we plot $h_{\text{m}}$ for
varying $U_{\downarrow }$, which exhibits the same behaviour as that
obtained from Eq.~(\ref{Umin}). The agreement between $h_{\text{m}}$ and
Eq.~(\ref{Umin}) signals that the picture of the effective interaction $U_{%
\text{eff}}$ works in a wide range of parameters, even inside the SF phase
where the density fluctuation is somehow enhanced. \

\subsection{Finite interaction bias: $U_{\uparrow }\!-\!U_{\downarrow }<\infty $}

\label{sec:results:finite}

\begin{figure}[tp]
\includegraphics[width=8.5cm]{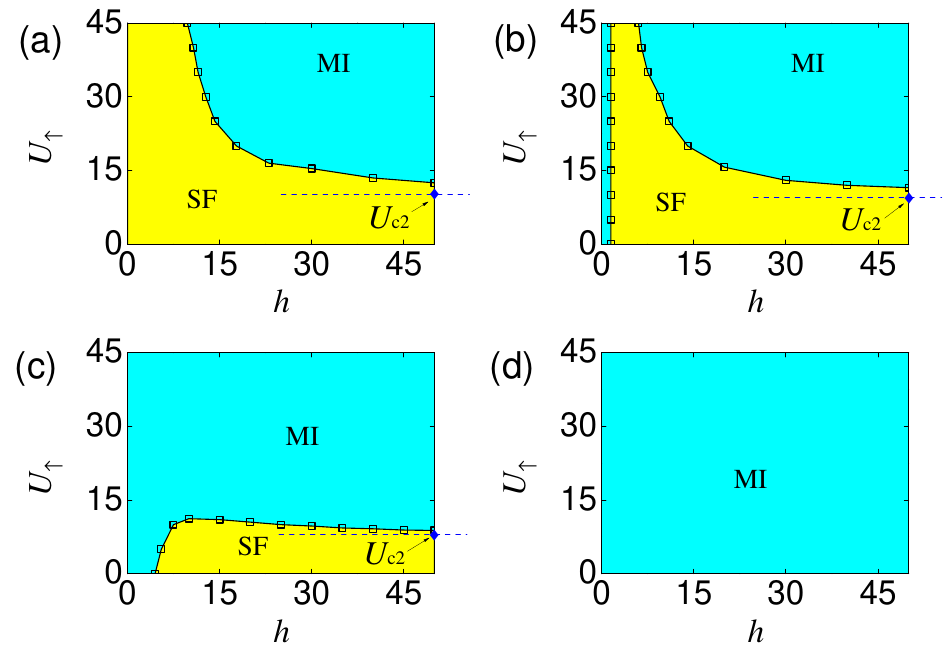}
\caption{The phase diagram in the $U_{\uparrow }-h$ plane with $\Delta =10$
and (a) $U_{\downarrow }=3.0$, (b) $U_{\downarrow }=3.6$, (c) $U_{\downarrow
}=5.0$, and (d) $U_{\downarrow }=15.0$.}
\label{Fig6}
\end{figure}

Having understood the physics of bosonic ladder under the infinite
interaction bias, we are now in the stage to explore the more general
parameter regime where the interactions of both chain of the ladder are
finite. Here we are particularly interested in the influence of finite
spin-up interaction on various quantum phases. By calculating the charge gap
$\delta _{L}$ with extrapolation to the thermodynamic limit, we obtain the
phase diagrams in the $U_{\uparrow }-h$ plane in Figs.~\ref{Fig6}(a)-(d)
with different $U_{\downarrow }$. As shown in Fig.~\ref{Fig6}(a), in which
the spin-down interaction is fixed as $U_{\downarrow }=3.0\ $($<$ $U_{\text{c}}$),
the SF region is confined by a smooth phase boudary, which extends up to $%
U_{\uparrow }\rightarrow \infty $\ and $h\rightarrow \infty $.\ A phase transition from the SF to
MI may occur when increasing $U_{\uparrow }$ ($h$) for some fixed $h$ ($%
U_{\uparrow }$). Increasing the spin-down interaction slightly larger than $%
U_{\text{c}}$, for example $U_{\downarrow }=3.6$, the MI can emerge for small $h$,
penetrating the SF region, as illustrated in Fig.~\ref{Fig6}(b).
Importantly, as $h$ approaches infinity, the spin-up interaction delimiting
different quantum phases decreases and saturates to some critical value $%
U_{\text{c2}}$.

In fact, through an analysis of the plaritonic modes, the MI-to-SF phase
boundary in the $h \rightarrow \infty $ limit can be derived as $U_{\uparrow
}+U_{\downarrow }=4U_{\text{c}}\approx 13.2$ (see Appendix B for details). Setting $%
U_{\uparrow }=U_{\downarrow }=U$, we immediately reproduce the result of the
symmetric case, i.e., $U=2U_{\text{c}}\approx 6.6$, obtained by the bosonization
method \cite{BHLtheory1}. Under this framework, the critical interaction $%
U_{\text{c2}}$ is straightforwardly written as
\begin{equation}
U_{\text{c2}}=4U_{\text{c}}-U_{\downarrow }.
\end{equation}%
As marked in Fig.~\ref{Fig6}(b) by blue dashed line, the critical interaction $%
U_{\text{c2}}$ defined above separates the phase diagram into two distinct
parameter regimes. For the $U_{\uparrow }>U_{\text{c2}}$ side, there exists the
interesting MI-to-SF-to-MI phase transition we explored in Subsection~\ref%
{sec:results:infinite}, whereas for $U_{\uparrow }<U_{\text{c2}}$, the MI-to-SF phase
transition can appear only once by monotonically varying $h$.

Adopting the description of the effective interaction introduced in
Subsection~\ref{sec:results:infinite}, we anticipate that if $U_{\downarrow
} $ is\ increased to be larger than $U_{\text{c1}}$, a finite upper bound of $%
U_{\uparrow }$, beyond which the SF phase disappears, can emerge. This is
confirmed by the phase diagram in Fig.~\ref{Fig6}(c), where we take $%
U_{\downarrow }=5$ ($>U_{\text{c1}}$=3.9). As expected, the SF phase is destroyed for $%
U_{\uparrow }\gtrsim 11$, in contrast to the behaviour shown in Figs.~\ref%
{Fig6}(a) and \ref{Fig6}(b). Increasing $U_{\downarrow }$ further such that $%
U_{\downarrow }>4U_{\text{c}}$, the critical interaction $U_{\text{c2}}$ touches zero,
meaning that the SF disappears, at least when $h$ is sufficiently large. The
phase diagram with $U_{\downarrow }=15$ ($>4U_{\text{c}}$) is plotted in Fig.~\ref%
{Fig6}(d), from which we find that the area of SF completely vanishes.

\begin{figure}[tp]
\includegraphics[width=8.5cm]{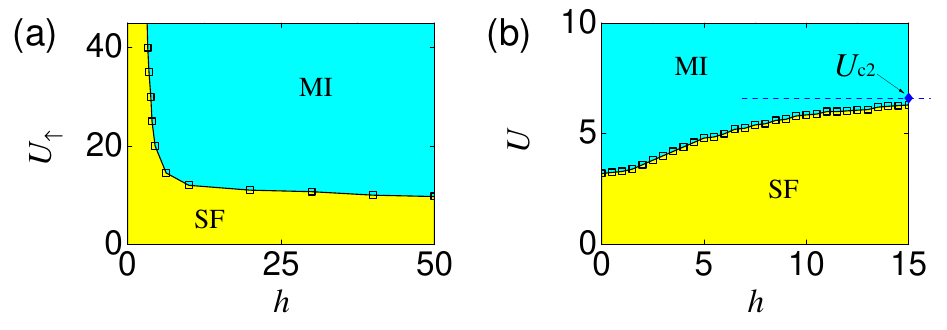}
\caption{(a) The phase diagram in the $U_{\uparrow }-h$ plane with $%
U_{\downarrow }=3.6$ and $\Delta =0$. (b) The phase diagram in the $U-h$
plane with $U=U_{\uparrow }=U_{\downarrow }$ and $\Delta =10$.}
\label{Fig7}
\end{figure}

Up to now, our focus are basically on the parameter regime where both the
interactions and potential energies of the two chains are asymmetric. The
individual effect incurred by one of the two asymmetric ingredients, i.e.,
either the interaction asymmetry or the potential energy asymmetry, has not
been elucidated. Here we complement this study by plotting two additional
phase diagrams, each of which has only one asymmetric ingredient. The phase
diagram in the $U_{\uparrow }-h$ plane with zero energy bias ($\Delta =0$) and
fixed spin-down interaction ($U_{\downarrow }=3.6$) is plotted in Fig.~\ref{Fig7}(a). The phase
diagram in this case shares the same structure with that in Fig.~\ref{Fig6}%
(a), albeit with shrunken SF area. It is also understood that no MI phase
can be found for sufficiently small $h$, contrasting the behaviour in Fig.~%
\ref{Fig6}(b), since a zero $\Delta $ always\ closes the charge gap for
lattices with non-integer filling at $h=0$. As shown in Fig.~\ref{Fig7}(b),
by requiring the interactions of the two chains to be equal, saying $%
U=U_{\uparrow }=U_{\downarrow }$, we map the phase diagram in the $U-h$
plane with finite energy bias $\Delta =10$. With the increase of $h$,\ \ the
critical interaction of the SF-to-MI transition monotonically increases,
asymptotically up to $U_{\text{c2}}\approx 6.6$, showing a distinct behaviour compared
to cases with asymmetric interactions. \

\section{Discussion and conclusion}

As mentioned in Sec.~\ref{sec:system}, the considered model can be directly
implemented with ultracold atoms inside optical lattices under various
experiment designs. For example, the bosonic ladder can be prepared by
growing optical superlattices, which forms a double-well structure along one
direction \cite{Current3,DOL1}. The tunnelling strength $h$, on-site
interactions $U_{\uparrow /\downarrow }$ and energy bias $\Delta $ can be
independently controlled by properly tuning the geometry of the optical
double well. Alternatively, one can employ the spin-dependent optical
lattices \cite{SOL1}, where atoms with different hyperfine states experience
different lattice potentials. In this scenario, $\Delta $ and $h$ are
respectively controlled by the detuning and Rabi frequency of an additional
coupling laser, and the on-site interactions $U_{\uparrow /\downarrow }$ can
be tuned via Feshbach resonances or the lattice depths experienced by atoms
with different spins. In spite of the research interests of the model in its
own right, our results offer beneficial insights into engineering effective
interactions on demand by dressing different atomic internal states \cite%
{Rydberg1,Rydberg4}. That said, our model constitutes only a subset of the
rich physics in the BH ladder with coupled chains, and many interesting
extensions are to be applied in the future. For example, with a ladder
structure, the hopping process of atoms may carry nontrivial peierls phases,
giving rise to synthetic gauge fields \cite{Gauge}. These gauge fields may
not only affect the MI-to-SF transitions dramatically \cite{ExtendBH4} but also induce various chiral currents \cite{Current1,Current2,Current3,Current4,Current5,Current6}. Another direction is to
fit the system into the grand-canonical description by introducing a tunable
chemical potential \cite{BHM1,Biased1}. This may provide new perspectives on the
magnetic or charge correlations in Mott lobes with different filling factors.

In conclusion, we have theoretically studied the ground-state properties of
the BH ladder with half filling in a biased configuration by using state-of-the-art DMRG
numerical methods. It is found that the interchain tunnelling can drive both
the MI-to-SF and SF-to-MI quantum\ phase transitions, depending on the value
of interactions. A reentrant quantum phase transition between MI and SF has
also been predicted by setting the on-site interactions to some intermediate
values. Under appropriate conditions, the model is shown to be amenable to
some analytical treatment, whose predictions about the phase boundary is
in great agreement with the numerical results. Armed with these knowledge,
we have mapped out the full phase diagram and characterized some critical
parameters separating the system into regimes with distinct phase behaviours.

\section*{Acknowledgments}
This work is supported by the National Key R\&D Program of China under Grant No. 2022YFA1404003, the National Natural Science Foundation of China (NSFC) under Grant No.~12004230, 12174233 and 12034012, the Research Project Supported by Shanxi Scholarship Council of China and Shanxi '1331KSC'.

\vbox{\vskip1cm} \appendix

\section{Plaritonic modes of the local Hamiltonian}

In this Appendix, we derive Eq.~(\ref{PE}) in the main text. To that end, we
rearrange the Hamiltonian~(\ref{BBH}) as%
\begin{equation}
\hat{H}=\sum_{j}\hat{H}_{L}^{(j)}-t\sum_{\left\langle i,j\right\rangle ,\sigma
}\hat{b}_{i,\sigma }^{\dag }\hat{b}_{j,\sigma }  \label{BBH2}
\end{equation}%
where%
\begin{eqnarray}
\hat{H}_{L}^{(j)} &=&-h(\hat{b}_{\uparrow }^{\dag }\hat{b}_{\downarrow }+\text{H.c.})+\Delta
(\hat{n}_{\uparrow }-\hat{n}_{\downarrow })  \notag \\
&&+\sum_{\sigma }\frac{U_{\sigma }}{2}\hat{n}_{\sigma }\left( \hat{n}_{\sigma }-1\right)
\label{LH}
\end{eqnarray}%
describes the local physics at lattice site $j$. Note that we have omitted
the subscript $j$ in the right hand side of Eq.~(\ref{LH}) for simplicity.
The Hamiltonian~(\ref{LH}) can be spanned by the Fock basis $\left\vert
n_{\downarrow },n_{\uparrow }\right\rangle $, where $\hat{n}_{\sigma }$ is the
occupation number for bosons with spin $\sigma $($=\uparrow ,\downarrow $).
In the $U_{\uparrow }\rightarrow \infty $ limit, a hardcore constraint on
the spin-up bosons can be imposed, meaning that we only need to retain the
states $\left\vert n,0\right\rangle $ and $\left\vert n-1,1\right\rangle $
with the total occupation $n=n_{\downarrow }+n_{\uparrow }$. The plaritonic
modes of Hamiltonian~(\ref{LH}) are therefore admixtures of $\left\vert
n,0\right\rangle $ and $\left\vert n-1,1\right\rangle $, and the
eigenenergies are readily diagonalized as

\begin{eqnarray}
\omega _{n}^{\pm } &=&\frac{U_{\downarrow }}{2}n(n-2)+\frac{1}{2}(\Delta
+U_{\downarrow })  \notag \\
&&\pm \frac{1}{2}\sqrt{(nU_{\downarrow }-U_{\downarrow }-\Delta )^{2}+4nh^{2}%
}
\end{eqnarray}

\section{Effective low-energy descritption in the $h \rightarrow \infty $ limit}

Here we provide an effective low-energy decryption of the model in the $%
h \rightarrow \infty $ limit. We first introduce two branches of quasi-modes $%
\hat{b}_{j,+}$ and $\hat{b}_{j,-}$ defined as%
\begin{equation}
\hat{b}_{j,+}=\frac{1}{\sqrt{2}}(\hat{b}_{j,\uparrow }+\hat{b}_{j,\downarrow })  \label{T1}
\end{equation}%
\begin{equation}
\hat{b}_{j,-}=\frac{1}{\sqrt{2}}(\hat{b}_{j,\uparrow }-\hat{b}_{j,\downarrow })  \label{T2}
\end{equation}%
Under the transformations of Eqs. (\ref{T1}) and (\ref{T2}), the Hamiltonian
(\ref{BBH}) is rewritten as
\begin{equation}
\hat{H}=\sum_{j}\hat{H}_{L}^{(j)}-t\sum_{\left\langle i,j\right\rangle }\left(
\hat{b}_{i,+}^{\dag }\hat{b}_{j,+}+\hat{b}_{i,-}^{\dag }\hat{b}_{j,-}\right)  \label{BBH3}
\end{equation}%
where the local Hamiltonian reads%
\begin{widetext}
\begin{eqnarray}
\hat{H}_{L}^{(j)} &=&(\frac{U_{\downarrow }}{8}+\frac{U_{\uparrow }}{8}%
)[(\hat{n}_{j,+}+\hat{n}_{j,-})^{2}+(\hat{b}_{j,+}^{\dag }\hat{b}_{j,-}+\hat{b}_{j,-}^{\dag
}\hat{b}_{j,+})^{2}-2(\hat{n}_{j,+}+\hat{n}_{j,-})]  \notag \\
&&+(\frac{U_{\downarrow }}{4}-\frac{U_{\uparrow }}{4})[(\hat{b}_{j,+}^{\dag
}\hat{b}_{j,-}+\hat{b}_{j,-}^{\dag }\hat{b}_{j,+})(\hat{n}_{j,+}+\hat{n}_{j,-}-1)]  \notag \\
&&+\frac{\Delta }{2}(\hat{n}_{j,+}+\hat{n}_{j,-}-\hat{b}_{j,+}^{\dag }\hat{b}_{j,-}-\hat{b}_{j,-}^{\dag
}\hat{b}_{j,+})+h(\hat{n}_{j,+}-\hat{n}_{j,-})  \label{LH2}
\end{eqnarray}%
\end{widetext}

It follows from Eq.~(\ref{LH2}) that the low-energy physics is dominated by
bosons on the \textquotedblleft $-$\textquotedblright\ plaritonic\ branch in
the $ h \rightarrow \infty $ limit. We thus anticipate an effective low-energy
theory which is purely described by field operators of the \textquotedblleft
$-$\textquotedblright\ plaritonic\ branch. The simplest way to achieve this
is to average the Hamiltonian~(\ref{BBH3}) with respect to the vacuum state
of the the \textquotedblleft $+$\textquotedblright\ plaritonic\ branch,
yielding

\begin{eqnarray}
\hat{H}_{\text{eff}} &=&\sum_{j}\left[ (\frac{\Delta }{2}-h)\hat{n}_{j,-}+\frac{\tilde{U}%
}{2}\hat{n}_{j,-}(\hat{n}_{j,-}-1)\right]  \notag \\
&&-t\sum_{\left\langle i,j\right\rangle }\hat{b}_{i,-}^{\dag }\hat{b}_{j,-}  \label{EffH}
\end{eqnarray}%
where $\tilde{U}=(U_{\downarrow }+U_{\uparrow })/4$. Notice that the
effective description in the Hamiltonian~(\ref{EffH}) becomes accurate when $%
h$ approaches infinity. More importantly, the Hamiltonian~(\ref{EffH})\
is written in the same form of the one-dimensional BH model with effective
on-site interaction $\tilde{U}$. It follows that the physics of our ladder
system in this limit can be effectively described by the one-dimensional BH
model with simple substitution of system parameters. Given this, the SF-to-MI
phase boundary is readily obtained as $\tilde{U}=U_{\text{c}}\approx 3.3$.

\end{document}